\documentclass[twocolumn,aps,prl]{revtex4}
\usepackage{graphicx}
\usepackage{bm}
\usepackage{amsmath}
\usepackage{amssymb}

\begin{document}
\title{Non-equilibrium thermodynamic study of magnetization dynamics
  in the presence of spin-transfer torque}

\author{Kazuhiko Seki and Hiroshi Imamura}
\affiliation{
  Nanotechnology Research Institute, National Institute of Advanced
  Industrial Science and Technology, 
  1-1-1 Umezono, Tsukuba, Ibaraki 305-8568, Japan
}

\begin{abstract}
  The dynamics of magnetization in the presence of
  spin-transfer torque was studied. We derived the equation for the motion of
  magnetization in the presence of a spin current by using the local
  equilibrium assumption in non-equilibrium thermodynamics. 
  We show that, in the resultant equation, the ratio of the Gilbert
  damping constant, $\alpha$, and the coefficient, $\beta$, of the
  current-induced torque, called non-adiabatic torque, depends on 
  the relaxation time of the fluctuating field $\tau_{c}$.
  The
equality $\alpha=\beta$ holds 
when $\tau_c$ is very short compared to the time scale of magnetization dynamics.
  We apply our theory to current-induced magnetization reversal
  in magnetic multilayers and show that the switching time is a
  decreasing function of $\tau_{c}$. 
\end{abstract}

\maketitle

Spin-transfer torque-induced magnetization dynamics such as 
current-induced magnetization reversal
\cite{slonczewski96,berger96,katine00}, domain wall motion \cite{klaui03}, 
and microwave generation \cite{kiselev03} have attracted a great deal of attention
because of their potential applications to future nano-spinelectronic devices.
In the absence of spin-transfer torque, magnetization dynamics is  
described by either the Landau-Lifshitz (LL) equation \cite{landau35} or the
Landau-Lifshitz-Gilbert (LLG) equation \cite{gilbert04}. It is known that the LL and
LLG equations become equivalent through rescaling of the
gyromagnetic ratio. 

\begin{figure}
\centerline{\includegraphics[width=0.95\columnwidth]{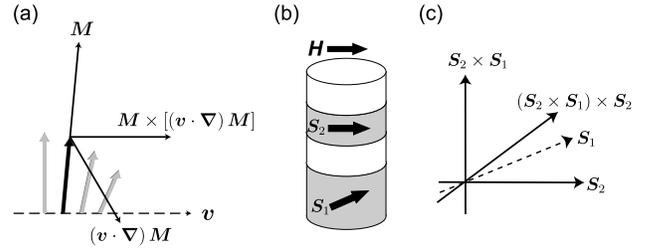}}
\caption{
  (a) The direction of the magnetization $\bm{M}$, the adiabatic
  contribution of spin-transfer torque,
  $\left(\bm{v}\cdot\bm{\nabla}\right)\bm{M}$, and the $\beta$ term,
  $\bm{M}\times\left[\left(\bm{v}\cdot\bm{\nabla}\right)\bm{M}\right]$,
  are shown.  The direction of the velocity $\bm{v}$ is indicated by
  the dotted arrow.
  (b)
  The magnetic multilayers, in  which the pinned
  and the free layers are separated by a
  nonmagnetic spacer layer are schematically shown.
  The magnetization vectors of the pinned and
  free layers are represented by $\bm{S}_{1}$ and $\bm{S}_{s}$, respectively. 
  The effective magnetic field to which $\bm{S}_{2}$ is subject is represented by $\bm{H}$.
 (c) The direction of the magnetization of the free layer,
  $\bm{S}_{2}$, the spin-transfer torque $(\bm{S}_{2}\times
  \bm{S}_{1})\times\bm{S}_{2}$, and the non-adiabatic torque,
  $\bm{S}_{2}\times \bm{S}_{1}$, are shown.  The direction of
  $\bm{S}_{1}$ is indicated by the
  dotted arrow.
  }
\label{fig:introduction}
\end{figure}

However, this is not the case in the presence of
spin-transfer torque.
For domain wall dynamics, the following LLG-type equation
has been studied by several groups \cite{thiaville05,barnes05,kohno06}:
\begin{align}
&\partial_{t} \langle \bm{M} \rangle    + 
\bm{v} \cdot \bm{\nabla}  \langle \bm{M} \rangle     
= \gamma  \bm{H} 
\times \langle \bm{M} \rangle  
\nonumber\\
&+ \frac{\alpha}{M} 
\langle \bm{M} \rangle \times  \partial_{t} \langle \bm{M} \rangle 
+ \frac{\beta}{M} 
 \langle \bm{M} \rangle \times
\left[ \left( \bm{v} \cdot \bm{\nabla} \right) 
\langle \bm{M} \rangle   \right],   
\label{eq:LLG_LLG_conv}
\end{align}
where $\bm{M}$ represents the magnetization, $\bm{v}$ is the velocity,
$\gamma$ is the gyromagnetic ratio and $\alpha$ is the Gilbert damping
constant.  The second term on the left-hand side represents the
adiabatic contribution of 
spin-transfer torque.  The first and the second terms on the
right-hand side are the torque due to the effective magnetic field $\bm{H}$ and
the Gilbert damping. The last term on the right-hand side of
Eq. \eqref{eq:LLG_LLG_conv} represents the current-induced torque, 
called ``non-adiabatic torque'' or simply the $\beta$ term. 
The directions of the adiabatic contribution of spin-transfer torque
and non-adiabatic torque are
shown in Fig. \ref{fig:introduction} (a).  

As shown by Thiaville {\it et al.},
the value of the coefficient $\beta$ strongly influences the motion of
the domain wall \cite{thiaville05}.
However, the value of the coefficient $\beta$ is still controversial, 
and different conclusions have been drawn from
different approaches \cite{barnes05,tserkovnyak06,tserkovnyak08,kohno06,xiao06,stiles07,duine07}.
For example, Barnes and Maekawa 
showed that the value of $\beta$ should be equal to that of the
Gilbert damping constant $\alpha$ to satisfy the requirement 
that the relaxation should cease at the minimum of electrostatic energy, even under particle flow.  
Kohno {\it et al.} performed microscopic calculations of spin
torques in disordered ferromagnets and showed that the $\alpha$ and
$\beta$ terms arise from the spin relaxation processes and that 
$\alpha\neq\beta$ in general \cite{kohno06}.
Tserkovnyak {\it et al.} \cite{tserkovnyak06} derived the $\beta$ term
using a quasiparticle approximation and showed that $\alpha=\beta$ 
within a self-consistent picture based on the local density approximation.

In the current-induced
magnetization dynamics in the magnetic multilayers shown in
Fig. \ref{fig:introduction} (b) \cite{ZhangLevy04,tulapurkar05,kubota07}, 
the non-adiabatic torque exerts a strong effect, and 
therefore affects the direct-current voltage of the spin torque diode, 
as shown in Refs. \cite{tulapurkar05,kubota07}. 
The magnetization dynamics of the free layer,
$\bm{S}_{2}$, has been studied by using the following LLG-type equation, 
\begin{align}
&\partial_{t} \bm{S}_2  -  
\frac{I}{e} g 
\hbar  
\left(\bm{S}_2 \!\times\! \bm{S}_1 \right) \times \bm{S}_2 
= \gamma  \bm{H} 
\times \bm{S}_2 
+ 
\frac{\alpha}{S_2} 
\bm{S}_2 \times \partial_{t} \bm{S}_2 
\nonumber\\
&
+ \eta I
\bm{S}_2 \times \bm{S}_1, 
\end{align}
where $I$ is the charge current density, $g$ is the amplitude of the spin torque 
introduced by Slonczewski \cite{slonczewski96}, $\hbar$ is the Dirac constant and 
$\eta$ represents the magnitude of the ``non-adiabatic torque'' 
which is sometimes called the field-like
torque \cite{tulapurkar05,kubota07}.

In this paper, we study the magnetization dynamics induced by 
spin-transfer torque in the framework of non-equilibrium thermodynamics.
We derive the equation of motion of the magnetization in the presence
of a spin current by using the local equilibrium assumption.
In the resultant equation, the Gilbert damping term and the
$\beta$ term are expressed as memory terms   
with the relaxation time of the fluctuating field $\tau_{c}$.
We show that the value of the coefficient $\beta$ is not equal to that of the Gilbert
damping constant $\alpha$ in general.  However, we also show that the
equality $\alpha=\beta$ holds if $\tau_{c} \ll 1/(\gamma H)$.
We apply our theory to the current-induced magnetization reversal
in magnetic multilayers and show that the switching time is
a decreasing function of $\tau_{c}$. 

Let us first briefly introduce the non-equilibrium
statistical theory of magnetization dynamics in the absence of spin
current \cite{miyazaki98}.
The LLG equation describing the motion of magnetization $\bm{M}$
under an effective magnetic field $\bm{H}$ is given by
\begin{equation}
  \partial_{ t} \bm{M}=\gamma\,\bm{H}\times\bm{M}+\frac{\alpha}{M}\bm{M}\times\partial_{ t} \bm{M}. 
  \label{eq:llg}
\end{equation}
The equivalent LL equation is
expressed as
\begin{equation}
 \partial_{ t} \bm{M}
  =\frac{\gamma}{1+\alpha^{2}}\bm{H}\times\bm{M}
  -\frac{\alpha \gamma}{M(1+\alpha^{2})}\bm{M}\times(\bm{M}\times\bm{H}).
  \label{eq:ll}
\end{equation}
The Langevin equations leading to Eqs. \eqref{eq:llg} and \eqref{eq:ll} 
by taking the ensemble average of magnetization $\bm{m}$, 
are 
\begin{align}
  &\partial_{ t} \bm{m} = \gamma\, \bm{H}_{\rm tot}\times \bm{m} 
  \label{eq:Langevin_1}
  \\
  &\partial_{ t} \delta \bm{H} = 
  - \frac{1}{\tau_c} \left( \delta \bm{H} - \chi_s \bm{m} \right) + \bm{R} (t),
  \label{eq:Langevin_2}
\end{align}
where the total magnetic field $\bm{H}_{\rm tot}$  is the sum of the
effective magnetic field $\bm{H}$ and the fluctuating magnetic field
$\delta \bm{H}$ and $\chi_s$ is the susceptibility of the local magnetic field 
induced at the position of the spin.  
According to Eq. \eqref{eq:Langevin_2} 
the fluctuating magnetic field $\delta \bm{H}$ relaxes toward the reaction
field $\chi_s \bm{m}$ with the relaxation time $\tau_c$. The
random field $\bm{R}(t)$ satisfies $\langle \bm{R} (t) \rangle =0$  and
the fluctuation-dissipation relation, 
$\langle R_i (t) R_j (t')  \rangle
= \frac{2}{\tau_c} \chi_s k_{\rm B} T \delta_{i,j} \delta (t-t')$, 
where $k_{\rm B}$ is the Boltzmann constant, $T$ is the temperature, 
$\langle \cdots \rangle$ denotes the ensemble average, 
and $i,j=1,2,3$ represents the Cartesian components.  It was shown that
Eqs. \eqref{eq:Langevin_1} and \eqref{eq:Langevin_2} lead to
Kawabata's extended Landau-Lifshitz equation \cite{kawabata72} derived
by the projection operator method \cite{miyazaki98}.
In the Markovian limit, {\it i.e.}, $\tau_c \ll 1/(\gamma H)$, we can obtain 
the LLG equation \eqref{eq:llg} and the corresponding LL equation
\eqref{eq:ll} with $\alpha=\gamma\tau_{c}\chi_{s}M$ \cite{miyazaki98}.

In order to consider the flow of spins, {\it i.e.}, spin current, we
introduce the positional dependence.  
Since we are interested in the average motion, 
it is convenient to introduce the mean velocity of the carrier, 
$\bm{v}$. 
The average magnetization, $\langle \bm{m}(\bm{x},t) \rangle$, is obtained 
by introducing the positional dependence and taking the ensemble average of Eq. \eqref{eq:Langevin_1}. 
In terms of the mean velocity, 
the ensemble average of the left-hand side of Eq. \eqref{eq:Langevin_1} leads to  
\begin{align}
\partial_{t} \langle \bm{m}  \rangle 
+ \left( \bm{v} \cdot \bm{\nabla} \right) \langle \bm{m}  \rangle . 
\end{align}
Assuming $\langle \delta \bm{H}  \times \bm{m} \rangle \approx 
 \langle \delta \bm{H} \rangle 
\times \langle \bm{m}  \rangle$, which is applicable when the thermal
fluctuation is small compared to the mean
value, 
we obtain 
\begin{eqnarray}
\partial_{t} \langle \bm{m}  \rangle 
+ \left( \bm{v} \cdot \bm{\nabla} \right) \langle \bm{m}  \rangle   \!=\! \gamma 
\langle \bm{H}_{\rm tot} (\bm{x},t) \rangle \times \langle \bm{m}(\bm{x},t) \rangle .  
\label{eq:eqmm} 
\end{eqnarray}
The mean magnetization density is expressed as $\langle \bm{M} (\bm{x},t) \rangle =  \rho (\bm{x},t) \langle \bm{m} (\bm{x},t) \rangle$, {\it i.e.},  
by the product of the scalar and vectorial components both of which depend on 
the position of the spin carrier at time $t$. 
The spin carrier density satisfies the continuity equation,
\begin{equation}
\partial_{t}\rho(\bm{x},t) + \bm{\nabla}\cdot \left( \bm{v} \rho(\bm{x},t) \right) =0. 
\label{eq:continuity}
\end{equation}
By multiplying the left-hand side of Eq. \eqref{eq:eqmm} by $\rho(\bm{x},t) $ 
and using the continuity equation \eqref{eq:continuity}, 
the closed expression for the mean magnetization is obtained as \cite{degroot}  
\begin{align}
\rho ( \partial_{t} \langle \bm{m} \rangle +  \bm{v} \cdot \bm{\nabla} \langle \bm{m} \rangle )
&=  \partial_{t} \rho\langle \bm{m} \rangle + 
\langle \bm{m} \rangle \bm{\nabla}\cdot \bm{v} \rho   
+\rho\bm{v} \cdot \bm{\nabla} \langle \bm{m} \rangle \nonumber \\
& =\partial_{t} \langle \bm{M} \rangle
+ \mbox{Div} \bm{v} \langle \bm{M} \rangle, 
\label{eq:eqml}
\end{align}
where 
$\mbox{Div} \bm{v} \langle \bm{M}  \rangle$ is defined by 
\begin{eqnarray}
\mbox{Div} \bm{v} \langle \bm{M} \rangle 
\!= \!\!\sum_{i=1}^3 
\frac{\partial \, v_i \langle \bm{M} \rangle }{\partial x_i}
\!= \langle \bm{M} \rangle(\bm{\nabla}\!\cdot\!\bm{v})
\!+\! (\bm{v}\!\cdot\!\bm{\nabla})\langle \bm{M} \rangle . 
\label{eq:defDiv}
\end{eqnarray}
By multiplying the right-hand side of Eq. \eqref{eq:eqmm} by $\rho(\bm{x},t) $ 
and using Eq. \eqref{eq:eqml}, we obtain 
\begin{eqnarray}
\partial_{ t} \langle \bm{M} \rangle + 
\mbox{Div} \bm{v} \langle \bm{M} \rangle = \gamma 
\left( \bm{H} + \langle \delta \bm{H} \rangle \right)
\times \langle \bm{M}  \rangle . 
\label{eq:constituent1}
\end{eqnarray}
Equation \eqref{eq:constituent1} takes the standard form of a 
time-evolution equation for extensive thermodynamical variables under
flow \cite{degroot}.  
The average of
Eq. \eqref{eq:Langevin_2} with the positional dependence is given by   
\begin{equation}
\partial_{ t}   \langle \delta \bm{H}  (\bm{x} ,t)\rangle 
  =
  -\frac{1}{\tau_c} \left[ \langle \delta \bm{H} (\bm{x} ,t)\rangle - \chi \langle
    \bm{M} (\bm{x} (t),t)\rangle \right] , 
  \label{eq:constituent2}
\end{equation}
where $\bm{x}(t)$ is the mean position at time $t$ of the spin carrier, which flows with velocity 
$\bm{v}=\partial_{ t}\bm{x} (t)$ and $\chi=\chi_s /\rho$ is assumed to be 
a constant independent of the position. 
Equations \eqref{eq:constituent1} and \eqref{eq:constituent2} constitute
the basis for the subsequent study of magnetization dynamics in 
the presence of spin-transfer torque.

The formal solution of Eq. \eqref{eq:constituent2} is expressed as
\begin{equation}
\langle \delta \bm{H}  (\bm{x} ,t) \rangle 
= \frac{\chi}{\tau_c} 
\int_{-\infty}^t \psi (t-t') \langle \bm{M} (\bm{x} (t'),t') \rangle\, dt' ,
\label{eq:dH_LLGa}
\end{equation}
where the memory kernel is given by $\psi (t) = \exp [-t/\tau_c] $.
Using partial integration, we obtain 
\begin{equation}
\langle \delta \bm{H}  (\bm{x},t)  \rangle 
= \chi\langle \bm{M} \rangle - \int_{-\infty}^{t}
\psi (t-t') \chi
\langle\dot{\bm{M}}(t')\rangle\,dt', 
\label{eq:dH_LLGb}
\end{equation}
where the explicit expression for $\dot{\bm{M}}(t)=\dot{\bm{M}}(\bm{x} (t),t)$ is 
given by the convective derivative,  
\begin{align}
\dot{\bm{M}}(t)&= 
 \partial_{t} \bm{M}  (\bm{x}(t), t) 
+\left( \bm{v} \cdot \bm{\nabla} \right) \bm{M} (\bm{x}(t), t) . 
\label{eq:convective}
\end{align} 
Substituting Eq. \eqref{eq:dH_LLGb} into Eq. \eqref{eq:constituent1}, 
we obtain the equation of motion for the mean magnetization density,
\begin{align}
&\partial_{t} \langle \bm{M} \rangle  + 
\mbox{Div} \bm{v} \langle \bm{M} \rangle   
= \gamma \bm{H} 
\times \langle \bm{M} \rangle  
\nonumber\\
&+ \gamma
\int_{-\infty}^t dt' \psi (t-t') \chi 
 \langle \bm{M} (t) \rangle \times \langle\dot{ \bm{M}} (t') \rangle .
\label{eq:eom_1}
\end{align}
Equation \eqref{eq:eom_1} supplemented by   
Eq. \eqref{eq:convective} is the {\em principal result} of this paper.

When the relaxation time of the fluctuating field, $\tau_{c}$, is very
short compared to the time scale of the magnetization dynamics,  
the memory kernel is decoupled and 
Eq. \eqref{eq:eom_1} can be written in the form of an LLG-type equation as
\begin{equation}
\partial_{t}  \langle \bm{M} \rangle  + 
\mbox{Div} \bm{v}  \langle \bm{M} \rangle  
= \gamma \bm{H}
\times  \langle \bm{M} \rangle  
+ \frac{\alpha}{M} 
 \langle \bm{M} \rangle \times \dot{\langle \bm{M}\rangle}, 
\label{eq:LLG_markov}
\end{equation}
where $\alpha = \gamma \tau_c \chi M$ is the Gilbert damping constant. 
Substituting the explicit form of the convective derivative, Eq. \eqref{eq:convective}, 
into Eq. \eqref{eq:LLG_markov} and using
Eq.\eqref{eq:defDiv} we obtain
the following LLG-type equation:  
\begin{align}
&\partial_{t} \langle \bm{M} \rangle  + 
\langle \bm{M} \rangle(\bm{\nabla}\!\cdot\!\bm{v})
+ (\bm{v}\!\cdot\!\bm{\nabla})\langle \bm{M} \rangle 
= \gamma \bm{H}   
\times \langle \bm{M} \rangle
\nonumber\\
&+ \frac{\alpha}{M} 
\langle \bm{M} \rangle  \times \partial_{t} \langle \bm{M} \rangle
+ \frac{\alpha}{M} 
\langle \bm{M} \rangle \times 
\left[ \left( \bm{v} \cdot \bm{\nabla} \right) 
\langle \bm{M} \rangle \right] .   
\label{eq:LLG_LLGa}
\end{align}
If $\bm{\nabla}\cdot\bm{v}=0$, Eq. \eqref{eq:LLG_LLGa} reduces to 
Eq. (14) of Ref. \cite{barnes05}, which is derived by replacing the
time derivative of magnetization $\partial_{t}\bm{M}$ on both sides of the LLG equation
\eqref{eq:llg} by the convective derivative $\partial_{t}\bm{M} +
\bm{v}\cdot\bm{\nabla}\cdot\bm{M}$.  
The term $\langle \bm{M} \rangle(\bm{\nabla}\!\cdot\!\bm{v})$ 
appears not on the right-hand side of
Eq. \eqref{eq:LLG_LLGa} but on the left-hand side, which means we cannot obtain
Eq. \eqref{eq:LLG_LLGa} using the same procedure
used in Ref. \cite{barnes05}.  
As shown in  Refs. \cite{barnes05,tserkovnyak07}, 
Eq. \eqref{eq:LLG_LLGa} with $\langle \bm{M} \rangle \left( \bm{\nabla}\cdot\bm{v} \right) = 0$ leads to 
a steady-state solution in the comoving frame, 
$\langle \bm{M} (t) \rangle   = \langle \bm{M}_0
(\bm{x}-\bm{v}t)\rangle $, where $\langle \bm{M}_0 (\bm{x}) \rangle $
denotes the stationary solution in the absence of domain wall motion. 
However,  if $\langle \bm{M} \rangle \left( \bm{\nabla}\!\cdot\!\bm{v} \right) \neq 0$, 
the steady-state solution may break the Galilean invariance.
The situation $\langle \bm{M} \rangle \left( \bm{\nabla}\cdot\bm{v} \right) \neq 0$ can be realized, for
example, in magnetic semiconductors \cite{ohno98,dietl06}, where the
spin carrier density is spatially inhomogeneous, {\it i.e.}, $\bm{\nabla}\rho\neq
0$.

The last term of Eq. \eqref{eq:LLG_LLGa} represents the non-adiabatic 
component of the current-induced torque, which is also known as the 
``$\beta$ term''.  
By comparing Eq. \eqref{eq:LLG_LLGa} with Eq. \eqref{eq:LLG_LLG_conv},  
one can see that the coefficient of the last term is equal to  
the Gilbert damping constant  $\alpha$.
However, Eq. \eqref{eq:LLG_LLGa} is valid 
when the relaxation time of the fluctuating field, $\tau_{c}$, is very
short compared to the time scale of the magnetization dynamics. 
It should be noted that the general form of the equation describing the magnetization dynamics 
is given by Eq. \eqref{eq:eom_1} where the last term on the
right-hand side is the origin of the $\alpha$ and $\beta$ terms. 
It is possible to project the torque represented by the memory function onto the direction
of the $\alpha$ and $\beta$ terms. 
This projection leads to  
$\alpha \neq \beta$ in general.

In order to observe the effect of $\tau_{c}$ on the magnetization
dynamics we applied our theory to the current-induced magnetization switching in 
the magnetic multilayer shown in Fig.\ref{fig:introduction} (b). 
We assumed that the fixed and free
layers are single-domain magnetic layers acting as a large spin 
characterized by the total magnetization vector defined as
$\bm{S}_i = \int dV \langle \bm{M}_i \rangle$, where $i=1(2)$ for the
fixed (free) layer and $\int dV$ denotes the volume integration over the
 fixed (free) layer. Both the magnetization vector of the fixed layer
 $\bm{S}_{1}$ and the effective magnetic field, $\bm{H}$, acting on the free layer lie  in the plane.

Integrating Eqs. \eqref{eq:constituent1} and \eqref{eq:constituent2}
over the volume of the free layer, we obtain the equations, 
\begin{align}
&\partial_{ t} \bm{S}_2  + 
\int dS \hat{\bm{n}} \cdot \bm{J}  = \gamma \left( \bm{H} + \langle  \delta \bm{H} \rangle \right) 
\times \bm{S}_2, 
\label{eq:Langevin1_c}\\
&\partial_{t} \langle \delta \bm{H} \rangle = 
-\frac{1}{\tau_c} \left( \langle \delta \bm{H} \rangle - \chi_V \bm{S}_2 \right) , 
\label{eq:Langevin2_c}
\end{align}
where $\bm{J}=\bm{v} \otimes \langle \bm{M} \rangle$  is the spin current tensor
$\int dS$ represents the surface integration over the free layer,  
$\hat{\bm{n}}$ is the unit normal vector of the surface,
and $\chi_V=\chi/V$ is defined by the volume of the free layer $V$. 

The same procedure used to derive Eq. \eqref{eq:eom_1} yields  
\begin{align}
&\partial_{t} \bm{S}_2  + 
\int dS \hat{\bm{n}} \cdot \bm{J}
= \gamma \bm{H}   
\times \bm{S}_2 
\nonumber\\
&+ \gamma
\int_{-\infty}^t dt' \psi (t-t') \chi_V  \bm{S}_2 (t) \times \partial_{t'} \bm{S}_2 (t'), 
\label{eq:dM_LLGb}
\end{align}
where $\psi (t) = \exp [-t/\tau_c] $. 

When the relaxation time of the fluctuating field is short compared to  
the time scale of magnetization dynamics, the LLG-type equation in the
presence of the spin-transfer torque  
is obtained as  
\begin{equation}
\partial_{t} \bm{S}_2  + 
\int dS \hat{\bm{n}} \cdot \bm{J}
= \gamma \bm{H} 
\times \bm{S}_2 
+ \frac{\alpha}{S_2} 
\bm{S}_2 \times \partial_{t} \bm{S}_2, 
\label{eq:LLG_LLGb}
\end{equation}
where $\alpha=\gamma \tau_c \chi_V S_2$.  
By introducing the conventional form of the spin-transfer torque
\cite{slonczewski96}, 
we obtain the following 
LLG-type equation:  
\begin{equation}
\partial_{t} \bm{S}_2  - 
\frac{I}{e} g \hbar  
\left(\bm{S}_2 \!\times\! \bm{S}_1 \right) \times \bm{S}_2 
\!=\! \gamma  \bm{H} 
\times \bm{S}_2 
+ \frac{\alpha}{S_2} 
\bm{S}_2 \times \partial_{t} \bm{S}_2.
\label{eq:LLG_LLGb_conv}
\end{equation}
However, Eq. \eqref{eq:LLG_LLGb_conv} is valid only when $\tau_c < 1/(\gamma H)$. 
As mentioned before, 
the torque represented by using the memory function generally has 
a component parallel to the non-adiabatic torque. 
In order to observe the effect of the non-adiabatic torque induced by the memory function 
on the magnetization dynamics, 
we performed numerical simulation using 
Eqs. \eqref{eq:Langevin1_c} and \eqref{eq:Langevin2_c}. 

For the simulation, we used the following conditions.
At the initial time of $t=0$, we assumed that 
the magnetization of the free layer is
aligned parallel to the effective magnetic field $\bm{H}$ and 
the angle between the magnetizations of the fixed and the free layers is 45$^{\circ}$.
This arrangement corresponds to the recent experiment on 
a magnetic tunnel junction system \cite{kubota07}. 
We also assumed that the fluctuation field has zero mean value at
$t=0$, {\it i.e.},
$\langle \delta \bm{H} (0) \rangle= \bm{0}$.

\begin{figure}
\centerline{\includegraphics[width=0.85\columnwidth]{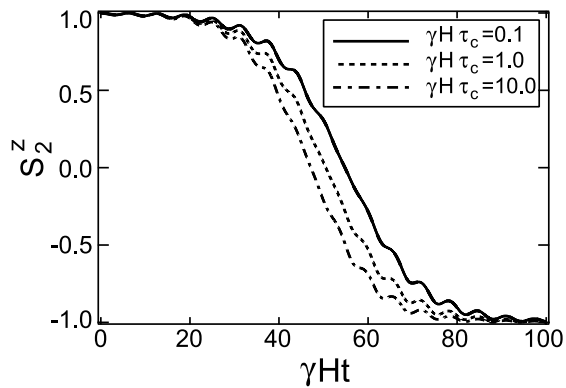}}
\caption{
  The $z$-component of the magnetization $\bm{S}_{2}$ is plotted as a
  function of time for various values of $\tau_{c}$. 
  The initial direction of the free layer is taken to lie in the direction of 
  the effective magnetic field, which is aligned to the $z$ axis. 
  The initial angle between
  $\bm{S}_{1}$ and $\bm{S}_{2}$ is taken to be 45$^{\circ}$.  
  The Gilbert damping constant $\alpha$ is 0.01. 
  }
\label{fig:switching}
\end{figure}

In Fig. \ref{fig:switching}, we plot the time dependence of the
$z$ component of the magnetization of the free layer, $\bm{S}_{2}$,
under the large-enough spin
current to flip the magnetization of the free layer,  $I g \hbar S_{2}^{2}
S_1/(e \alpha \gamma H) = -10$.  The value of $\tau_c$ is varied 
while the value of $\alpha=0.01$ is maintained. The solid, dotted, and
dot-dashed lines correspond to $\gamma H\tau_c=0.1, 1.0$, and 10.0,
respectively. 
As shown in Fig. \ref{fig:switching}, the time 
required for the magnetization of the free layer to flip decreases
with increasing $\tau_c$, which can be understood by considering the 
non-adiabatic torque induced by the spin current.  
The non-adiabatic torque induced by the spin current is obtained
by projecting the torque given by the last term of
Eq. \eqref{eq:dM_LLGb} onto the direction of
$\bm{S}_{2}\times \bm{S}_{1}$, 
which results in the positive contribution to the
spin-flip motion of $\bm{S}_{2}$.
Since the last term of Eq. \eqref{eq:dM_LLGb} includes a memory function,
the non-adiabatic torque induced by the spin current increases with
increasing $\tau_c$.  Therefore, the time 
required for $\bm{S}_{2}$ to flip decreases with
increasing $\tau_c$.  
For $\gamma H \tau_c>10$ we observe no further
decrease of the time required for $\bm{S}_{2}$ to flip because the
memory function is an integral of the vector $\bm{S}_2 (t)
\times \partial_{t'} \bm{S}_2 (t')$ and the contributions from 
the memory at $t-t' \gg 1/(\gamma H) $ is eliminated.

In conclusion, we derived the equation for the motion of
magnetization in the presence of a spin current by using the local
equilibrium assumption in non-equilibrium thermodynamics. 
We demonstrated that the value of the coefficient $\beta$ is not equal to that of the Gilbert
damping constant $\alpha$ in general.  However, we also show that the
equality $\alpha=\beta$ holds if $\tau_{c} \ll 1/(\gamma H)$.
We then applied our theory to current-induced magnetization reversal
in magnetic multilayers and showed that the switching time is
a decreasing function of $\tau_{c}$. 

The authors would like to acknowledge the valuable discussions they had with 
S.E. Barnes, S. Maekawa, P. M. Levy, K. Kitahara, K. Matsushita, J. Sato and T. Taniguchi. This work was
supported by NEDO.


\end{document}